\documentclass[pra,twocolumn,longbibliography,nofootinbib]{revtex4-1}
\usepackage[utf8]{inputenc}
\usepackage{soul,color}

\usepackage{bm,amsfonts,amsmath,amssymb}
\usepackage{dcolumn,xfrac}
\usepackage{easyReview}
\usepackage{natbib}
\usepackage[colorlinks]  
{hyperref}

\hypersetup{
    colorlinks=true,
    linkcolor=blue,
    citecolor=red,
    filecolor=magenta,
    urlcolor=magenta,
}

\newcommand{\Eref}[1]{Eq.\,(\ref{#1})}

\begin{document}
\title{Sensing: Equation One}

\author{Dmitry Budker}
\affiliation{Johannes Gutenberg-Universit{\"a}t Mainz, 55128 Mainz, Germany}
 \affiliation{Helmholtz-Institut, GSI Helmholtzzentrum f{\"u}r Schwerionenforschung, 55128 Mainz, Germany}
\affiliation{Department of Physics, University of California, Berkeley, California 94720, USA}

\author{Mikhail~G.~Kozlov}

\affiliation{Petersburg Nuclear Physics Institute of NRC ``Kurchatov
Institute'', Gatchina 188300, Russia}

\affiliation{St.~Petersburg Electrotechnical University
``LETI'', Prof. Popov Str. 5, 197376 St.~Petersburg}

\date{\today}

\begin{abstract}
Spin projection noise sets a limit for the sensitivity of spin-based magnetometers and experiments searching for parity- and time-reversal-invariance-violating dipole moments. The limit is described by a simple equation that appears to have universal applicability.
\end{abstract}

\maketitle

\section{Introduction}
Spin-based devices such as atomic magnetometers (see, for example, \cite{Budker2007optical,Budker2013optical}) are among the most sensitive sensors finding numerous uses in both practical \cite{grosz2017high,fu2020sensitive} and fundamental-science \cite{Budker2013optical} applications.

A common starting point for the analysis of the sensitivity of a spin-based magnetometer is an equation that assumes that the experiment has overcome technical sources of noise and imperfections and is limited solely by the fundamental spin-projection noise. The equation reads:
\begin{equation}
    \delta B \approx \frac{\hbar}{g\mu_0\sqrt{2J}}\left(\frac{\Gamma}{NT}\right)^{1/2} .
    \label{Eq:One}
\end{equation}
Here $\delta B$ is the uncertainty in the determination of magnetic field $B$, $\hbar$ is the reduced Planck constant, $g$ is the Land\'{e} factor for the particular type of spin system used, $\mu_0$ is the Bohr magneton, $J$ is the spin of the system (or its total angular momentum), $\Gamma$ is the spin-relaxation rate, $N$ is the number of spins partaking in the measurement, and $T$ is the total time the spins evolve in the presence of the magnetic field. We will assume that this time is approximately equal to the total measurement time. In this article, we deliberately ignore various numerical factors of order unity for a maximally transparent and general discussion. 

Equation \eqref{Eq:One} can be written in various forms, and has appeared at the beginning of magnetometry papers often enough, so nowadays, it is frequently referred to as ``Equation One'' in scientific presentations.

In this article, we recall an intuitive derivation of \Eref{Eq:One} and discuss how it appears in contexts, where it may not be initially obvious that this equation should apply, using the example of linear resonant Faraday rotation (the Macaluso-Corbino effect \cite{macaluso1898sopra}, see also \cite{budker2002resonant}). 

The analysis of sensitivity of spin-based magnetometers can be directly extended to experiments that look for permanent electric dipole moments (EDM) of atoms, molecules, and nucleons \cite{khriplovich2012cp}. Such moments violate the fundamental symmetries of parity and time-reversal invariance, and are of key importance as probes for physics beyond the standard model.

A nonzero EDM $\bm{d}=d\,(\bm{J}/J)$ leads to linear Stark effect that is an analog of the linear Zeeman effect due to the magnetic dipole moment $\bm{\mu}=g\mu_0 \bm{J}$ of a particle. Thus, the sensitivity to an EDM can be written as
\begin{equation}\tag{1a}
    \delta d \approx \frac{\hbar\sqrt{J/2}}{E}  \left(\frac{\Gamma}{NT}\right)^{1/2} ,
    \label{Eq:One_EDM}
\end{equation}
where $E$ is the electric field applied to the spins (or, in the case of atomic and molecular EDM, the effective electric field, see \cite{khriplovich2012cp}).

While modern EDM experiments operate at the limits predicted by \Eref{Eq:One_EDM}\,\cite{panda2019attaining}, it is tempting to think of ways to improve the measurements beyond these limitations. For example, one direction of active research over the past decades has been application of spin-squeezing and quantum entanglement of the spins. Such quantum methods are, in principle, capable of, for example, replacing the $N^{-1/2}$ scaling of Eqs.\,\eqref{Eq:One} and \eqref{Eq:One_EDM} with the $N^{-1}$ (Heisenberg) scaling. 
%
%
Unfortunately, spin-squeezed and entangled states are generally more susceptible to decoherence, so measurement schemes with squeezing or entanglement often end up offering only minor (if any) advantages \cite{auzinsh2004can}. We will not further discuss such approaches here. Instead, we consider the sensitivity of optical-rotation magnetometers, which is also relevant to EDM-induced optical-rotation experiments \cite{sushkov1978parity,barkov1988amplification} that have recently attracted renewed interest \cite{chubukov2019optical}.

\section{Sensing with spins}

\begin{figure*}[!htpb]\centering
	\includegraphics[width=1.0\columnwidth]{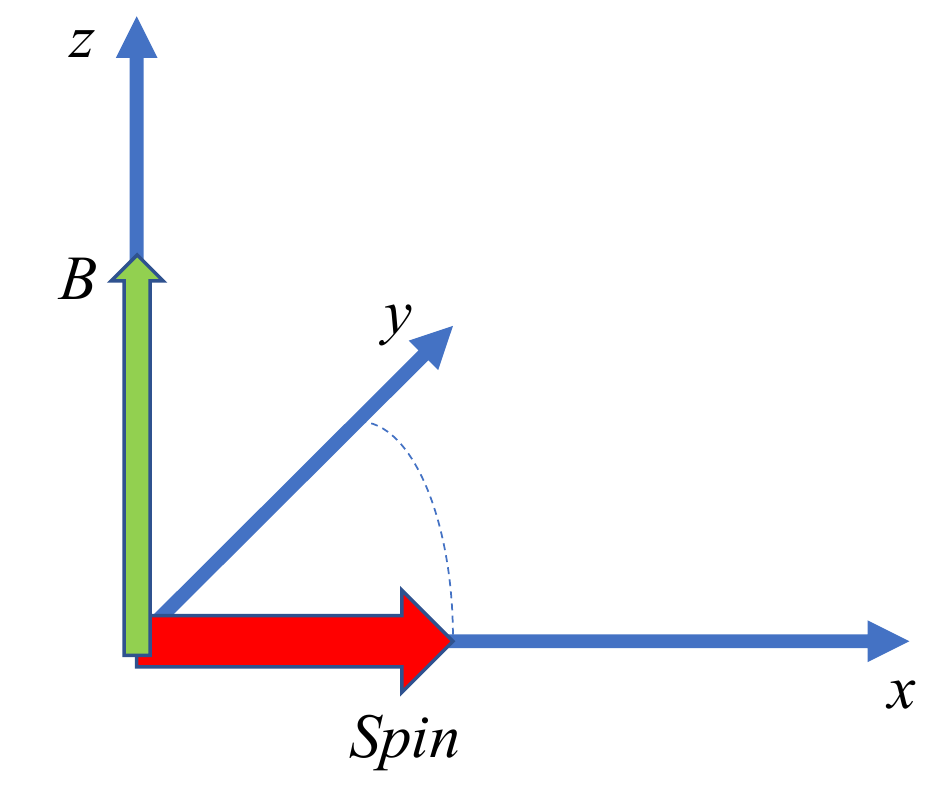}
	\caption{Schematic of an idealized spin-based measurement of a magnetic field assumed to be applied along $z$. A spin is prepared in the eigenstate with maximal projection along $x$ and is allowed to evolve in the presence of the magnetic field for time $T_1$. After that, a measurement of the spin projection along $y$ is performed, from which the value of the magnetic field is inferred. Such a ``pump-precession-probe'' scheme is ubiquitous in sensing, see, for example, \cite{Budker2007optical,Budker2013optical}.}
    \label{Fig:SensEqOneFig}
\end{figure*}
Let us consider the following idealized magnetometric experiment (Fig.\,\ref{Fig:SensEqOneFig}) whose aim is to measure a magnetic field with magnitude $B$ applied along $z$; other components of the magnetic field are assumed to be zero. A single spin, let us say, $s=1/2$ with a Land\'{e} factor $g$, is prepared in the eigenstate corresponding to maximum projection ($m_s=1/2$ in this case) along the Cartesian $x$ axis. Once the spin is prepared, it is allowed to freely precess for a time $T_1$ under the influence of the magnetic field. The magnetic field, without the loss of generality of our arguments, may be assumed weak, such that the spin-precession angle is small: $\varphi=g\mu_0 B T_1/\hbar\ll 1$. 
At the end of the precession, the projection of the spin on the $y$ axis is measured with 100\% efficiency.

According to quantum mechanics, the result of the measurement of the $y$ projection of the spin can be either $+1/2$ or $-1/2$. For $B=0$, the probability of these outcomes is equal, but in the presence of a magnetic field, one of  the two outcomes is slightly more probable: $P_{\pm 1/2}=(1\pm\varphi)/2$.

Suppose that we do the experiment and measure the $y$ projection of the spin to be +1/2. What can we say about the magnetic field? Not too much: perhaps we can claim that counter-clockwise rotation of the spin during the precession time is more probable than clockwise rotation. However, the uncertainty on the precession angle is complete: $\delta \varphi \approx 1$. However, if we repeat the experiment many times so that the total precession time is $T\gg T_1$, we can reduce the uncertainty by a factor of $(T/T_1)^{1/2}$. Thus, we can estimate the sensitivity of our cumulative magnetic measurement as 
\begin{equation}
    \delta B \approx \frac{\hbar}{g\mu_0}\frac{1}{T_1}\left(\frac{T_1}{T}\right)^{1/2}.
\end{equation}
If the measurements are carried out with $N$ independent spins, the uncertainty is further reduced by $N^{1/2}$, and we arrive at Eq.\eqref{Eq:One} for the case of $J=1/2$ if we identify $1/T_1$ with effective spin-relaxation rate $\Gamma$. A more formal derivation for an arbitrary $J$ is presented in the Appendix. 

\section{Linear Faraday rotation (the Macaluso-Corbino effect)}

Let us now consider a seemingly unrelated problem of linear (meaning low-light-power) Macaluso-Corbino effect (see \cite{budker2002resonant}) on an isolated atomic transition between the ground state with total angular momentum $J=0$ and and excited state with $J'=1$. A linearly polarized light beam passes through a sample of $N$ atoms, and the angle of polarization  rotation is detected in the transmitted light. For simplicity, we assume that the upper-state radiative width is $\Gamma$ and there are no other widths in the problem, for instance, we ignore the Doppler width. Suppose there is only a magnetic field applied in the direction of propagation of the light. This setup is used as magnetometer: the value of the magnetic field can be deduced from the polarization-rotation angle.

Let us  first  assume resonant light. The rotation angle can be estimated (see,  for example, \cite{budker2002resonant,budker2008atomic}) as:
\begin{align}
\label{Eq:LinFarRotAngle}    
\phi \approx \left(\frac{g\mu_0 B}{\hbar\Gamma}\right) \times \frac{l}{l_0}\,,
\end{align}
where $l$ and $l_0$ are the length of the interaction region and the absorption length, respectively. Here we assume that $|g\mu_0 B/\Gamma|\ll 1$. Note that, according to \Eref{Eq:LinFarRotAngle}, at small magnetic fields, the optical-rotation angle is linear in the field. (At higher fields, when $|g\mu_0 B/\Gamma|\approx 1$, the linearity breaks down, and the maximum rotation angle is on the order of a radian per absorption length.)

An important assumption in \Eref{Eq:LinFarRotAngle} is that light intensity is sufficiently low, such that the optical properties of the medium (absorption length and optical rotation per unit magnetic field per unit length, the Verdet constant) are independent of the light intensity.
 
The magnetometry protocol based on this scheme consist of measuring the rotation angle for a time $T$ and backing off the value of the field using \Eref{Eq:LinFarRotAngle}.

A fundamental noise source in such a measurement is the photon shot noise, which, for an ideal polarimeter, is given by (see \cite{budker2008atomic}):
\begin{align}
\label{Eq:PolarimNoise}    
\delta\phi = \frac{1}{2\sqrt{N_\mathrm{phot}}} \,,
\end{align}
where $N_\mathrm{phot}$ is the number of photons transmitted trough the atomic sample.

Let us now optimize the sensitivity of the magnetometer assuming that we can arbitrarily adjust the number of atoms and light intensity. The first instinct is to increase $N$ in order to increase $l/l_0$ and the rotation angle \eqref{Eq:LinFarRotAngle}. However, if the sample comprises many absorption lengths, only a small fraction of the photons will be transmitted, and this will increase the polarimetric noise, according to \Eref{Eq:PolarimNoise}. An optimization in $N$ (or equivalently, $l_0$) yields the optimal value of $l/l_0=2$.

Next, let us optimize the light intensity assuming that $N$ is fixed. We want as much light as possible (in order to reduce the polarimetric shot noise). However, we can only go up to power levels that correspond to optical saturation where we begin to bleach the atoms. In this regime, rotation angle drops with further increase of the light intensity and one no longer gains an advantage due to the increase in the flux of transmitted photons. Thus, we will not consider the high-light-intensity regime and assume saturation parameter of order unity for further discussion. Macaluso-Corbino rotation in the case of arbitrary light intensities and optical densities of the sample is discussed in \cite{auzinsh2010optically} and references therein.
 
For a unity saturation parameter, an atom absorbs a photon in a time $\approx 1/\Gamma$, so the number of absorbed photons per unit time is approximately: $N\Gamma.$

Now recall that we have chosen $l/l_0 = 2$, so  the number  of transmitted photons is on the same order as the number of absorbed photons, which is
$N\Gamma T\,,$ and the statistical uncertainty in the measurement of the rotation angle \eqref{Eq:LinFarRotAngle} is:
\begin{equation}
\delta \phi \approx  \frac{1}{\sqrt{N\Gamma T}}.\label{Eq:RotAngUncertainty}
\end{equation}
 
Combining equations \eqref{Eq:LinFarRotAngle} and \eqref{Eq:RotAngUncertainty}, we arrive at the magnetometer sensitivity of
\begin{align}\label{EqOneAgain}
\delta B \approx \frac{\hbar}{g\mu_0}\left(\frac{\Gamma}{N T}\right)^{1/2}\,,  
\end{align} 
which is identical \Eref{Eq:One}. 

This result is remarkable in that we used the assumption that the noise in the Macaluso-Corbino magnetometer comes from the photon shot noise and did not explicitly consider the noise due to atomic spins. Nevertheless, optimization of the sensitivity fixed the photon intensity and, correspondingly, the photon shot noise to a value that is related to the atom number $N$. 

One might wonder, could it help (assuming fixed $N$) to detune the light frequency far away from resonance, which allows to increase the light intensity before the onset of bleaching. Far away from resonance, the Macaluso-Corbino rotation falls as $1/\Delta^2$ \cite{budker2002resonant,Budker2007optical}, where $\Delta$ is the amount of detuning, while the light intensity corresponding to the onset of bleaching increases by the same factor \cite{auzinsh2010optically}. As a result, the signal-to-noise for far-detuned light is worse. 

Another tempting idea is to use an optical cavity to enhance the optical rotation. While such approach may, indeed, have practical advantages \cite{Bougas2012,Visschers2020}, an analysis of the fundamental sensitivity (that we do not reproduce here) brings us back to \Eref{Eq:One}. In conclusion of this section, we mention that the sensitivity of linear optical rotation experiments with warm atoms is still lower by typically several orders of magnitude due to Doppler broadening of the optical line reducing the peak rotation by a factor of $\approx\Gamma/\Gamma_D$, where $\Gamma_D$ is Doppler width, while also requiring a larger $N$ to achieve $l/l_0 = 2$.
 
\section{Discussion}

The emergence of the same limit in different settings suggest its universality. This is not entirely surprising. After all, the underlying mechanism in both magnetometer types is the Zeeman effect. How much the magnetic field affects the system depends on the factor $g\mu_0$ and the effective time the system is allowed to evolve under the action of the field before it relaxes (or is perturbed by the measurement process). Combined with the square-root scaling of the noise as a function of the spin or photon number, this results in \Eref{Eq:One}.  

An important question is whether it is possible to design an optimized magnetometer where, for example, the number of spins $N$ with intrinsic relaxation rate $\Gamma$ is fixed and one can otherwise perform a measurement with a sensitivity that would outperform \Eref{Eq:One}?

Apart from using squeezing as mentioned in the Introduction, there are various other ideas that have been put forward. One such idea is to replace a ``passive'' measurement with one involving a laser or a maser, see, for example, \cite{inoue2013nuclear}. The motivation is that such devices display linewidths that are orders of magnitude narrower than the ``pasive'' relaxation rates $\Gamma$ and the hope inspired by the scaling with $\Gamma$ of the sensitivity in equations \eqref{Eq:One} and \eqref{Eq:One_EDM} that this will translate in a better magnetic or EDM sensitivity. However, so far, such devices have not surpassed the sensitivity limits of \eqref{Eq:One} or \eqref{Eq:One_EDM}. We are not aware of a comprehensive theoretical analysis that would prove that equations \eqref{Eq:One} or \eqref{Eq:One_EDM} represent optimal sensitivity for active devices, however, this is so for all cases known to us insofar. 

Finally, the sensitivity limits discussed in this article are not the only case of seemingly universal limits related to sensing. Such a limit appears to exist for the energy resolution of magnetic sensors that tends to be limited by $\hbar$, as discussed in a detailed review paper \cite{Mitchell2020EnRes}.

\section*{Acknowledgement}
This work was stimulated by discussions with D. V. Chubukov, L. V. Skripnikov, A. N. Petrov, V. N. Kutuzov, L. N. Labzowsky, and L. Bougas and 
supported in part by the DFG Project ID 390831469:  EXC 2118 (PRISMA+ Cluster of Excellence).

\bibliographystyle{apsrev}
\bibliography{SensingEqOne}

\appendix
\section{Formal derivation of Equation (1)}
\label{Appendix:Formal_Derivation}

Let us consider quantum system with total angular momentum $\hbar\bm{J}$ and magnetic moment $\bm{\mu}=g\mu_0\bm{J}$ in the magnetic field $\bm{B}=B \hat{\bm{z}}$. We prepare the system in the state with maximum projection of the angular momentum on the axis ${x}$ and detect projection of the angular momentum on the axis ${y}$ (see Fig.\ \ref{Fig:SensEqOneFig}). 

Let us designate initial wave function as:
\begin{align}
    \label{Eq:Psi0}
    \Psi_0 &= |J,J\rangle_x\,.
\end{align}
Evolution in the magnetic field is described by the Hamiltonian
\begin{align}
    \label{Eq:Ham}
    \hat{H}&=-\bm{\mu}\bm{B} = -g\mu_0 B J_z\,,
\end{align}
After the sufficiently short time $\delta t$ the wave function becomes $\Psi_0+\delta \Psi$, where
\begin{align}
    \label{Eq:dPsi}
    i\hbar\delta\Psi 
    &= \hat{H}\Psi_0 \delta t
    = (-g\mu_0 B\delta t)\, J_z|J,J\rangle_x
    \nonumber \\
    &= (-g\mu_0 B\delta t)\, i \sqrt{J/2}|J,J-1\rangle_x
    \,,
\end{align}
where we use the fact that operator $J_z$ acts on the state $|J,J\rangle_x$ in the same way as operator $J_y$ acts on the state $|J,J\rangle_z$. Remembering that $J_y=-i/2(J_+-J_-)$ we obtain the answer given above. 

The signal, which we measure is the projection on the axis $y$:
\begin{align}\label{Eq:Sig}
    S &= |\langle\Psi_0 +\delta\Psi |J_y|\Psi_0 +\delta\Psi\rangle|
    = \frac{g\mu_0 B}{\hbar}\, \delta t\, J
    \,.
\end{align}
The noise is given by the dispersion of the measurement:
\begin{align}\label{Eq:Noise}
     \sigma &= \sqrt{\langle J_y^2\rangle-\langle J_y\rangle^2}
    \approx \sqrt{\langle \Psi_0| J_y^2|\Psi_0\rangle}=\sqrt{J/2}\,,
\end{align}
where we neglect all contributions, which depend on $\delta\Psi$. The $\sqrt{J/2}$ factor can be also derived using the vector model, where angular-momentum vectors are represented by ``cones'', see, for example, \cite{auzinsh2010optically}.

The signal \eqref{Eq:Sig} linearly grows with evolution time $\delta t$, but this time is limited by the relaxation rate: $\delta t\le \Gamma^{-1}$.
Thus, for a single measurement we have the signal-to-noise ratio:
\begin{align}\label{Eq:SN1}
    \frac{S}{\sigma} &= \frac{g\mu_0 B}{\hbar} \sqrt{2J}\,\Gamma^{-1}\,.
\end{align}
If we do the measurement for the ensemble of $N$ atoms and repeat it $N_r$ times we can improve signal-to-noise ratio by a factor $\sqrt{NN_r}$. For the $N_r$ measurements one needs at least the time $T=N_r\Gamma^{-1}$, therefore the final signal-to-noise ratio is   
\begin{align}\label{Eq:SNfin}
    \frac{S}{\sigma} &= \frac{g\mu_0 B}{\hbar} \sqrt{2J}\,\left(\frac{NT}{\Gamma}\right)^{1/2}\,.
\end{align}
The smallest detectable magnetic field corresponds to $S/\sigma=1$, which gives us \Eref{Eq:One}.
\end{document}